\documentclass[conference,11pt,a4paper]{IEEEtran}

\usepackage{amsmath,amsxtra,amssymb,amsthm,latexsym,amscd,amsfonts}

\makeatletter
\def\ps@IEEEtitlepagestyle{\pagestyle{empty}}
\makeatother

\ifCLASSINFOpdf
  \usepackage[pdftex]{graphicx}
  \DeclareGraphicsExtensions{.pdf,.jpeg,.png,.jpg}
\else
  \usepackage[dvips]{graphicx}
  \DeclareGraphicsExtensions{.eps}
\fi

\usepackage{array}
\usepackage[font=footnotesize]{subfig}

\usepackage{booktabs}
\usepackage{tabularx}
\usepackage{multirow}
\usepackage{marvosym}
\usepackage{cite}
\usepackage{orcidlink}
\usepackage{hyperref}
\usepackage[all]{hypcap}
\usepackage{url}
\hypersetup{
  pageanchor=false,
  linktoc=all,
  colorlinks=false,
  pdfborder={0 0 1},
  linkbordercolor={0 1 0},
  citebordercolor={0 1 0},
  urlbordercolor={0 1 0},
  bookmarks=true,
  bookmarksopen=true
}

\newcolumntype{L}[1]{>{\raggedright\arraybackslash}p{#1}}
\renewcommand{\abstractname}{Abstract}
\renewcommand{\IEEEkeywordsname}{Keywords}

\usepackage{microtype}
\microtypesetup{protrusion=false}
\makeatletter
\renewcommand{\@IEEEsectpunct}{\quad}
\makeatother
\usepackage[none]{hyphenat}
\begin{document}
\renewcommand{\abstractname}{Abstract}
\renewcommand{\IEEEkeywordsname}{Keywords}
\pagenumbering{gobble}

\title{Design and Evaluation of a\\ Controlled Post-Alert Incident
Orchestration and\\ Response Subsystem Using a Rule Engine and\\
a Local Large Language Model}

\author{\IEEEauthorblockN{Hoang-Lam Huynh\textsuperscript{1,2}, Quoc-Cuong Tang\textsuperscript{1,2}, Van-Tri Phan\textsuperscript{2,(\Letter)}, and Khuong Nguyen-An\textsuperscript{3,4,(\Letter)}\,\orcidlink{0000-0002-9910-6387}}
\IEEEauthorblockA{\textsuperscript{1}Ho Chi Minh City College of Transport, Ho Chi Minh City, Vietnam\\\textsuperscript{2}Academy of Cryptography Techniques, Ho Chi Minh City Campus, Vietnam\\
\textsuperscript{3}Faculty of Computer Science and Engineering,\\ Ho Chi Minh City University of Technology (HCMUT),\\ 268 Ly Thuong Kiet Street, Dien Hong Ward, Ho Chi Minh City, Vietnam\\
\textsuperscript{4} Vietnam National University Ho Chi Minh City, Linh Xuan Ward, Ho Chi Minh City, Vietnam\\
Emails: \texttt{hhlam@hcmct.edu.vn}, \texttt{tqcuong@hcmct.edu.vn}\\\texttt{phanvantri@actvn.edu.vn}, \texttt{nakhuong@hcmut.edu.vn}}}

\maketitle
\begin{abstract}
This paper presents a controlled post-alert incident orchestration and response subsystem for educational information systems. The architecture separates deterministic classification, contextual analysis, human approval, and technical execution. A Rule Engine determines severity and selects the playbook, while Static RAG and a local large language model provide advisory content under Validator, Guardrail, Output Sanitizer, and Safe Fallback controls. Experiments begin after simulated alerts are stored in Elasticsearch. The Rule Engine matched the predefined routing matrix in all 30 boundary cases. The Durable Queue completed 100 events without duplicate tasks, new failed tasks, or unintended firewall rules. An eight-alert contention experiment preserved the configured limit of one active model request, and 30 sequential measurements showed an overall mean post-alert processing time of approximately 33 seconds. The results demonstrate functional correctness, traceability, controlled recovery, and bounded model integration within the evaluated laboratory scope.
\end{abstract}

\begin{IEEEkeywords}
post-alert incident response, security orchestration,
Rule Engine, Static RAG, local LLM.
\end{IEEEkeywords}

\IEEEpeerreviewmaketitle

\section{INTRODUCTION}\label{sec:introduction}

The increasing use of online learning platforms, academic information systems, shared computer laboratories, and network services has produced a growing volume of security events in educational environments. Security monitoring tools can identify suspicious activities and generate alerts, but the subsequent investigation and response processes often remain dependent on administrators. Each alert must be reviewed, classified, correlated with operational context, and mapped to an appropriate response. When the alert volume increases, this manual workflow may lead to delayed handling, inconsistent decisions, or accidental disruption of legitimate users.

Automating post alert response is more difficult than directly converting an alert into a firewall rule. Educational networks may include dynamically assigned addresses, shared gateways, network address translation, and legitimate traffic bursts that resemble malicious behavior. A response action applied to an incorrectly identified address may affect multiple users or interrupt an academic service. Therefore, a practical response subsystem must verify protected addresses, prevent duplicate processing, distinguish between risk levels, record each decision, and allow administrators to retain control over actions that may affect normal operations. It must also handle interruptions safely because replaying an action after a partial execution may create duplicate or inconsistent firewall rules.

Large language models can support security operations by summarizing alerts, explaining attack behavior, and generating contextual recommendations. However, their outputs may be incomplete, unsupported, inconsistent, or unsuitable for direct execution. For this reason, the language model in this study is not allowed to determine alert severity, select response playbooks, or execute network actions. These decisions remain under deterministic rules and operational policies. The model is used only as an advisory component, and its output is processed through static retrieval augmented generation, schema validation, guardrails, output sanitization, and safe fallback mechanisms before being stored or presented to an administrator.

This paper presents a controlled post alert incident orchestration and response subsystem that combines Elasticsearch, a SQLite durable queue, a rule engine, static retrieval augmented generation, a local large language model, human approval, and a response engine. The proposed architecture separates four responsibilities. The rule engine determines the severity level and response playbook. The local language model provides contextual analysis. The administrator approves cases that require human intervention. The response engine performs controlled actions with whitelist verification, time limits, rollback support, and audit logging. This separation prevents the language model from changing technical decisions while still allowing it to assist incident analysis.

The main contributions of this study are threefold. First, it proposes a post alert response architecture that separates deterministic decision making, language model based analysis, human approval, and technical execution. Second, it integrates safety mechanisms including whitelist verification, durable event processing, human in the loop control, output validation, sanitization, safe fallback, time limited enforcement, rollback, and recovery without playbook replay. Third, it evaluates the proposed subsystem through functional tests, interruption recovery scenarios, a queue stress experiment with one hundred events, and a model resource contention experiment with eight alerts in a controlled laboratory environment.

The evaluation begins after alerts are stored in Elasticsearch and therefore measures only the post-alert processing path, not intrusion-detection accuracy or end-to-end network detection latency.

The remainder of this paper is organized as follows. Section~\ref{sec:related_work} reviews related work on security information and event management, security orchestration and response, and controlled use of language models in security operations. Section~\ref{sec:proposed_architecture} describes the proposed architecture and processing workflow. Section~\ref{sec:experimental_setup} presents the experimental environment and evaluation procedure. Section~\ref{sec:results} discusses the functional, recovery, queue, and model contention results. Section~\ref{sec:limitations} outlines the limitations of the current implementation. Section~\ref{sec:conclusion} concludes the paper and identifies directions for future development.

\section{RELATED WORK}\label{sec:related_work}

\subsection{Security Monitoring and Incident Response}

Security monitoring systems collect and organize events from network devices, servers, applications, and security controls. Log management guidance emphasizes the need to preserve event records, support investigation, and maintain sufficient information for later analysis \cite{ref1}. Elasticsearch and related data processing components are commonly used to store, search, and visualize large volumes of security events \cite{ref2}. However, the availability of centralized event data does not by itself determine how an alert should be prioritized or which response action should be performed.

Incident response guidance describes a structured process that includes preparation, detection, analysis, containment, recovery, and post incident review \cite{ref3}. Response playbooks further support consistent handling by defining expected actions for common incident scenarios \cite{ref4}. These approaches provide an operational foundation, but their implementation still requires local policies, approval mechanisms, and safeguards against inappropriate actions. In environments with shared addresses, dynamic address assignment, or limited security personnel, directly translating an alert into a blocking rule may create operational risk.

Security orchestration, automation, and response platforms address part of this problem by coordinating tools and automating predefined workflows. Previous work has demonstrated the use of orchestration engines to connect security components and execute response activities \cite{ref5}. Nevertheless, an automated workflow must still distinguish between alert classification, administrative approval, and technical execution. A playbook selection does not necessarily mean that the associated action is safe to perform, particularly when the available evidence is incomplete or when the target environment contains protected services.

Empirical studies of SOAR platforms have also examined analyst interaction, orchestration effectiveness, and the operational trade-offs of security automation \cite{ref7}.

\subsection{Rule Based Orchestration and Human Control}

Rule based processing provides a transparent method for mapping observable alert attributes to predefined severity levels and response playbooks. Compared with probabilistic decisions, deterministic rules are easier to inspect, reproduce, and test at boundary values. They are also suitable for enforcing organizational policies, such as preventing actions against whitelisted addresses or requiring approval before a medium risk event is contained.

Human involvement remains necessary when a response may interrupt legitimate services or when the available evidence does not justify autonomous enforcement. In a human in the loop workflow, the system prepares the incident context and recommended action, while an administrator retains the authority to approve or reject the response. The technical execution result should also be recorded separately from the administrative decision. An approved action may still fail because of a firewall error, and such a failure must not be interpreted as a rejected administrative decision.

\subsection{Controlled Use of Large Language Models}

Recent literature has examined the use of large language models across cybersecurity tasks, including analyst assistance and security operations \cite{ref8}. Large language models can assist security analysts by reorganizing technical evidence, explaining alert context, and generating investigation recommendations. Their ability to process natural language makes them useful for presenting fragmented event attributes in a form that is easier for an administrator to review. However, model generated content may be incomplete, inconsistent, unsupported by the available evidence, or unsuitable for direct execution.

For this reason, the language model in the proposed subsystem is not treated as a decision authority. Alert severity and playbook selection are established before the model is invoked. The model receives the alert evidence together with contextual knowledge and produces advisory content only. The final output is checked by a parser and a field level validator. Unsupported values are rejected or replaced with information obtained from deterministic components. When the model is unavailable or its response cannot be used, the workflow continues with a safe fallback rather than allowing the model failure to change the technical decision.

The subsystem uses a deterministic static retrieval mechanism instead of a vector database. Each normalized alert category is mapped to a versioned knowledge record containing a MITRE ATT and CK reference, a standard operating procedure, investigation steps, mitigation information, and rollback conditions. The same category and knowledge base version therefore produce the same retrieved context. This design favors reproducibility and auditability, although its coverage depends on manually maintained category mappings. In the thesis, Static RAG is explicitly implemented through exact category matching rather than embedding based semantic retrieval. Retrieval-augmented generation has also been investigated as a mechanism for grounding cybersecurity-oriented language models in external domain knowledge \cite{ref9}.

MITRE ATT and CK is used to standardize the description of behaviors and provide contextual references for investigation \cite{ref6}. The mapping does not prove that a specific technique has succeeded, and it does not determine the severity or response playbook. Instead, the mapping supports reporting, contextual analysis, and knowledge retrieval. This distinction prevents a reference identifier from being interpreted as direct evidence of a confirmed attack.

\subsection{Research Gap and Positioning of This Study}

Existing guidance on log management, incident response, and response playbooks emphasizes evidence preservation, structured incident handling, and consistent operational procedures \cite{ref1}, \cite{ref3}, \cite{ref4}. Research on security orchestration, automation, and response further demonstrates how heterogeneous security components can be coordinated through predefined workflows \cite{ref5}. However, these approaches do not by themselves specify how authority should be divided among deterministic classification, language-model-assisted analysis, human approval, and technical execution when a response action may affect shared users or educational services.

The inclusion of a local large language model introduces an additional control problem. Model-generated analysis may help present alert context and investigation recommendations, but it may also contain incomplete, unsupported, or operationally inappropriate content. MITRE ATT\&CK provides standardized contextual references \cite{ref6}, but an ATT\&CK mapping does not determine alert severity, select a response playbook, authorize an action, or establish that generated advice is safe. A controlled post-alert workflow therefore requires explicit validation, Guardrail replacement, output sanitization, deterministic fallback, and clear restrictions on model authority.

This study is positioned at the intersection of deterministic orchestration, controlled language-model assistance, and operational safety. The proposed subsystem does not treat the language model as an autonomous decision maker. Severity and playbook selection remain under the Rule Engine, administrative authorization is recorded separately from technical execution, and network actions are restricted to the Response Engine. The architecture additionally integrates durable queueing, duplicate prevention, two-stage whitelist verification, human-in-the-loop approval, time-limited enforcement, rollback, audit recording, and recovery without automatic playbook replay.

To make this positioning experimentally testable, the study evaluates not only functional and recovery behavior but also the necessity of model invocation. A deterministic Rule Engine and Static RAG configuration is compared with a Qwen-assisted configuration using the same structured alert cases, knowledge records, and output requirements. This comparison examines whether local model inference provides measurable value when predefined rules and versioned knowledge records are already available. Accordingly, the contribution of this study is an integrated and evaluated control architecture for post-alert incident handling rather than a new intrusion-detection model or a general-purpose SOAR platform.

Recent work has further explored context-aware security orchestration architectures that combine automation with controlled decision processes \cite{ref10}.

\section{PROPOSED ARCHITECTURE AND METHOD}\label{sec:proposed_architecture}

\subsection{Overall Architecture}

\begin{figure}[htbp]
    \centering
    \includegraphics[
        width=0.78\columnwidth,
        keepaspectratio
    ]{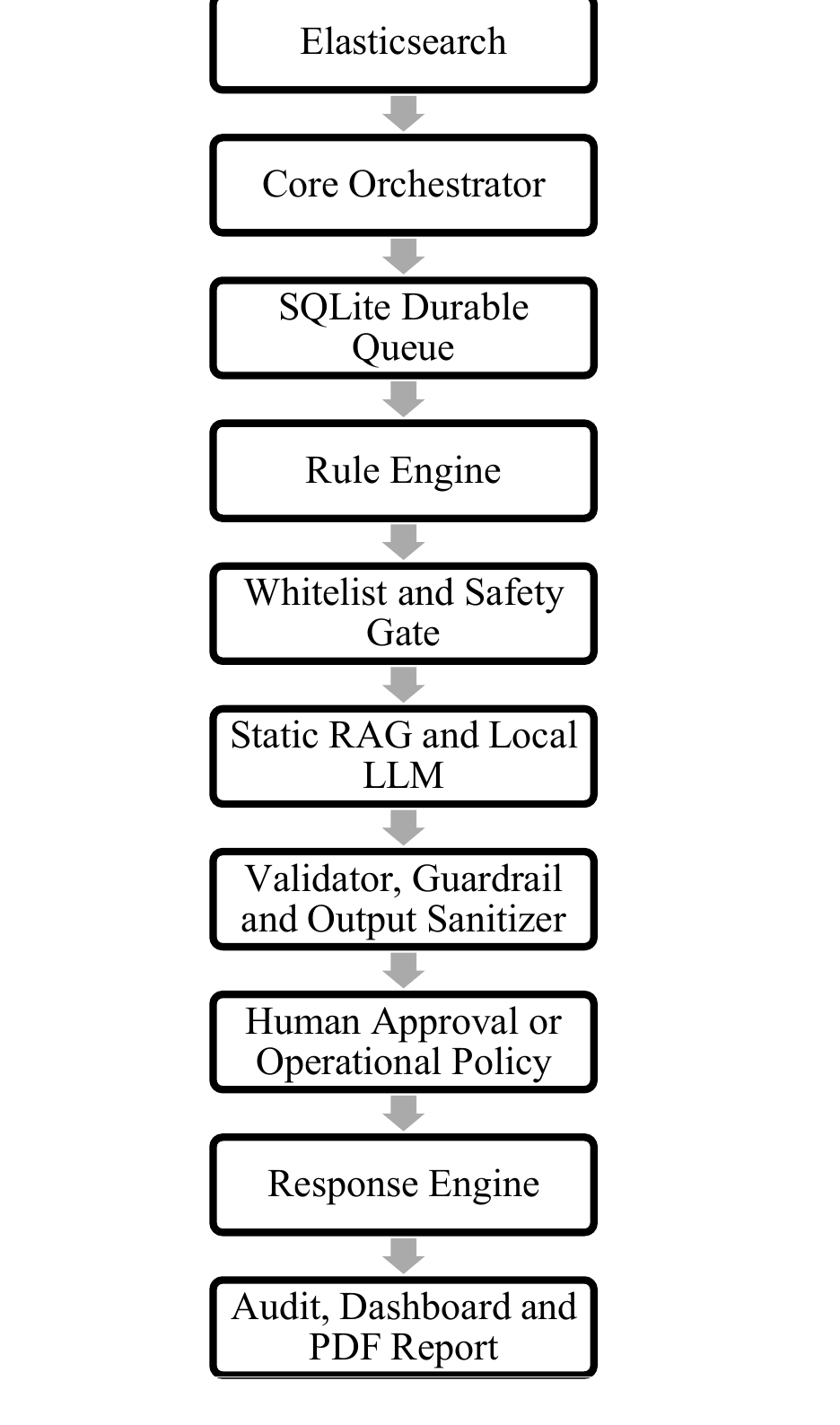}
    \caption{Overall architecture of the proposed post-alert incident orchestration and response subsystem.}
    \label{fig:overall_architecture}
    \pdfbookmark[2]{Figure \thefigure: Overall architecture}{bookmark:fig:overall_architecture}
\end{figure}

Figure~\ref{fig:overall_architecture} presents the overall architecture
of the proposed subsystem, which operates after a security alert has
been stored in Elasticsearch. It does not capture network packets, generate alerts, or determine the detection accuracy of an intrusion detection system. Its responsibility begins with retrieving an existing alert and continues through normalization, queue management, deterministic classification, contextual analysis, administrative approval, controlled response, and audit recording.

The architecture consists of five functional layers. The first layer contains Elasticsearch and the alert ingestion mechanism. Elasticsearch preserves the input alert documents, while the Core Orchestrator retrieves new documents and normalizes their relevant fields. The second layer provides durable processing and deterministic decisions through a SQLite based event queue, the Rule Engine, Whitelist and Safety Gate checks, and the Watchlist. The third layer performs contextual analysis using Static RAG, a local large language model, a JSON parser, a Validator and Guardrail, and an Output Sanitizer. The fourth layer manages administrative interaction and technical execution through the Core API, human approval, and the Response Engine. The final layer stores evidence and supports observation through SQLite audit records, the Dashboard, notifications, and PDF reports.

The Core Orchestrator acts as the central control point. External interfaces such as the Dashboard and Telegram do not directly modify firewall rules. Administrative decisions are submitted to the Core, which verifies the incident identifier, current state, policy, and whitelist status before forwarding an approved action to the Response Engine. Similarly, the local language model cannot change the severity, selected playbook, time limit, or firewall state. This allocation of authority reduces direct coupling between advisory, administrative, and execution components. The thesis architecture likewise centralizes control in the Core and requires administrative requests to be validated before reaching the Response Engine.

\subsection{Alert Ingestion and Durable Processing}

The processing workflow starts when the Core Orchestrator retrieves alert documents from Elasticsearch. Point in Time queries and the search\_after mechanism are used together with a checkpoint and a configurable lookback period. The lookback interval allows the Core to revisit a limited portion of the previous search range and reduce the possibility of missing documents that arrive late.

Each retrieved document is normalized into a common internal representation. The principal fields include the Elasticsearch document identifier, event time, source address, alert category, observed frequency, and the original input data. Alert categories are mapped into the supported groups of denial of service, brute force, exploit, and suspicious scan. Invalid records are retained as processing evidence but are not forwarded to the Response Engine.

After normalization, the event is inserted into the SQLite durable queue. The Elasticsearch document identifier is subject to a uniqueness constraint, which prevents the same input document from creating multiple queue tasks. Temporary failures may be retried within a configured limit. Tasks that contain invalid data or exceed their retry limit are moved to the DEAD state for later review. The separation between ingestion and processing allows the subsystem to control worker concurrency, preserve task state, retry temporary failures, and recover unfinished work after a Core interruption. These mechanisms and the unique es\_id constraint are defined in the thesis processing flow.

The subsystem uses three storage layers. Elasticsearch stores the original alert documents. SQLite stores queue state, incidents, operational policies, whitelist and watchlist entries, response rules, metrics, and audit records. The local file system stores configuration files, the versioned Static RAG knowledge base, generated reports, and experimental artifacts. This separation preserves the original alert data while isolating the operational state created by the Core.

\subsection{Deterministic Classification and Risk Based Routing}

The Rule Engine is invoked before the language model. It uses normalized alert attributes and predefined thresholds to determine the severity and corresponding playbook. These values are treated as authoritative throughout the remainder of the workflow. The language model may explain the decision or suggest investigation steps, but it cannot modify either value.

Before an event proceeds to a response branch, the subsystem evaluates its source address through the Whitelist and Safety Gate. The whitelist is checked at two stages. The first check occurs during initial processing to prevent protected addresses from entering an unsafe response path. The second check occurs immediately before technical execution so that an address added to the whitelist after initial classification is still protected.

\begin{figure}[!t]
    \centering
    \includegraphics[
        width=0.78\columnwidth,
        keepaspectratio,
        trim=0 30 0 0,
        clip
    ]{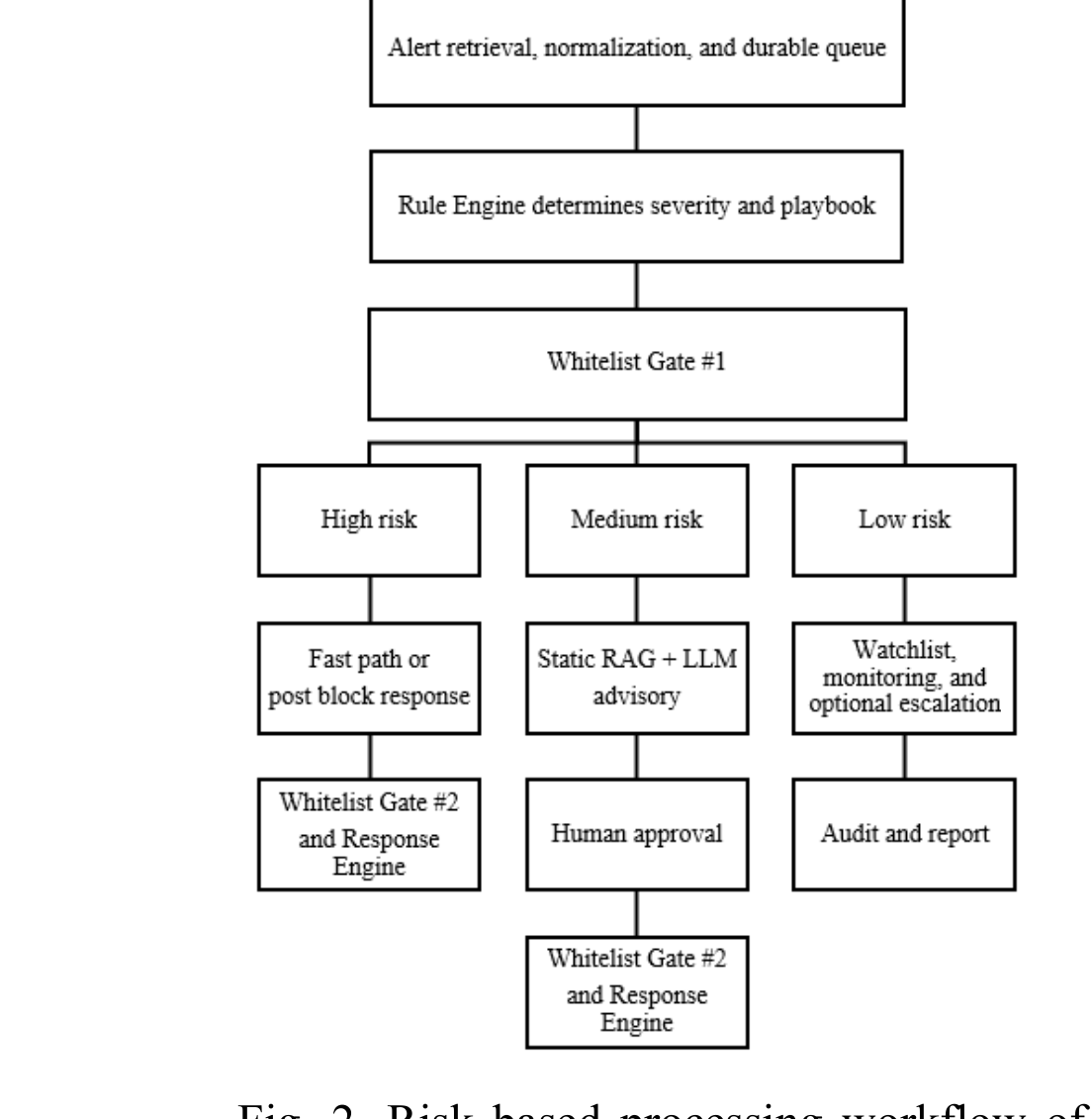}
    \caption{Risk based processing workflow of the proposed subsystem.}
    \label{fig:risk_workflow}
    \pdfbookmark[2]{Figure \thefigure: Risk based processing workflow}
    {bookmark:fig:risk_workflow}
    \vspace{-8 pt}
    
\end{figure}

Events are routed according to their determined risk level. A high risk event may enter a conditional fast path when the active policy and safety requirements permit containment. A medium risk event is analyzed and submitted for human approval before any enforcement action. A low risk event is added to the Watchlist and remains under observation. Repeated activity may increase its strike count and trigger controlled escalation. The architecture therefore uses different processing modes rather than
applying one response sequence to every alert, as illustrated in
Fig.~\ref{fig:risk_workflow}.

The authority of each component is summarized in Table~\ref{tab:responsibilities}.

\begin{table}[!htbp]
\caption{RESPONSIBILITIES AND CONTROL BOUNDARIES}\pdfbookmark[2]{Table \thetable: Responsibilities and control boundaries}{bookmark:tab:responsibilities}
\label{tab:responsibilities}
\centering\scriptsize
\renewcommand{\arraystretch}{0.70}
\setlength{\tabcolsep}{2pt}
\begin{tabularx}{\columnwidth}{@{}L{0.24\columnwidth}L{0.34\columnwidth}X@{}}
\toprule
\textbf{Component} & \textbf{Primary responsibility} & \textbf{Restricted action}\\
\midrule
Elasticsearch & Store and provide input alerts & Does not determine severity or playbook\\
Durable Queue & Preserve task identity and processing state & Does not authorize a response\\
Rule Engine & Determine severity and playbook & Does not execute firewall actions\\
Static RAG & Provide versioned contextual knowledge & Does not change technical decisions\\
Local LLM & Generate advisory analysis & Does not approve or execute responses\\
Validator and Guardrail & Validate and replace unsupported output & Does not select a playbook\\
Administrator & Approve or reject controlled actions & Does not directly modify the firewall through the interface\\
Response Engine & Execute an approved and validated action & Does not independently classify alerts\\
\bottomrule
\end{tabularx}
\end{table}

\subsection{Controlled Contextual Analysis}

After the Rule Engine has produced a severity and playbook, the subsystem retrieves contextual information from a versioned Static RAG knowledge base. Retrieval is performed through exact matching against the normalized alert category. Each knowledge record may contain a knowledge identifier, a MITRE ATT and CK reference, a risk description, a standard operating procedure, investigation steps, mitigation guidance, and rollback conditions.

Static RAG does not use embeddings, a vector database, or semantic search. This decision provides deterministic retrieval: the same category and knowledge base version produce the same contextual record. The approach is easier to reproduce and audit, although it depends on manually maintained category mappings and does not automatically cover new alert types.

The retrieved context and alert evidence are supplied to the local language model through one of three processing modes. The high risk mode supports post response analysis after a containment action. The medium risk mode prepares advisory information before human approval. The low risk mode supports monitoring and further investigation. In every mode, severity, playbook, time limits, and firewall state are treated as read only values.

The model response is not used directly. A JSON parser first checks whether the expected structure can be obtained. The Validator and Guardrail then inspect field types, allowed values, evidence consistency, and component authority. Unsupported or invalid fields are replaced with values from the Rule Engine or Static RAG. When the model times out, is unavailable, or returns an unusable response, the Core switches to a safe Static RAG fallback without changing the severity or playbook.

An Output Sanitizer is applied after validation and before the result is written to a database or displayed through the Dashboard, report, or notification channel. Its role is to remove unsafe markup and control characters, constrain excessive content, neutralize unsafe resource identifiers, and normalize unexpected structures. Metadata describing the validator, sanitizer, knowledge version, and replaced fields is stored with the incident for later review. The thesis specifies that the LLM is advisory only and that all output must pass parsing and validation before use.

\subsection{Controlled Execution, Recovery, and Audit}

The Response Engine is the only component permitted to apply technical response actions. Before execution, it rechecks the whitelist, validates the source address, confirms the current incident and rule states, and evaluates the active operating mode. The engine supports dry run operation, blocking, rate limiting, time limited rules, and rollback. A selected playbook therefore represents an intended response rather than proof that a firewall action has been successfully completed.

Human decisions and technical results are stored separately. An administrator may approve a response, while the subsequent firewall operation may still fail. In such a case, the approval state remains approved, whereas the execution state records the technical failure and its error details. This distinction prevents the system from confusing an accepted administrative decision with a successful enforcement result.

The subsystem also applies a non replay recovery policy to incidents interrupted during the EXECUTING state. When the Core restarts, it compares the stored incident with the observed firewall state. If the expected rule can be verified, the incident is completed without executing the playbook again. When the rule cannot be verified, the incident is moved to SAFETY\_HOLD for manual review. The Core does not automatically replay the response because doing so could create duplicate rules or repeat an external effect whose previous outcome is uncertain.

Throughout the workflow, the subsystem records the input evidence, Rule Engine decision, whitelist results, Static RAG context, language model state, fields replaced by the Guardrail, administrative decision, execution outcome, errors, and processing times. These records support the Dashboard, PDF reporting, experimental measurements, and audit review. A completed queue task only indicates that the Core has finished its assigned workflow; it does not independently prove that a firewall rule has been created. The actual enforcement outcome must be established from the Response Engine status and firewall evidence.

\section{EXPERIMENTAL SETUP AND EVALUATION METHOD}\label{sec:experimental_setup}

\subsection{Experimental Scope and Environment}

The experimental environment was designed to evaluate the functional behavior of the proposed post alert subsystem under controlled and repeatable conditions. The evaluation begins when an alert document is available in Elasticsearch. It does not include packet capture, attack traffic generation, intrusion detection model training, or measurement of detection accuracy. A simulator was used to generate structured alert documents and write them directly to Elasticsearch.

The experimental deployment included the alert simulator, Elasticsearch, the Core Orchestrator, SQLite, Ollama, the Dashboard, the Response Engine, and a laboratory firewall. The components were deployed on a single physical computer through local processes, Windows Subsystem for Linux, or Docker containers. Consequently, the reported results represent the behavior of the prototype under a specific hardware and software configuration rather than a distributed production deployment.

\subsection{Alert Data and System Configuration}

The simulator generated controlled alert documents with fields compatible with the main concepts of the Elastic Common Schema. The principal input fields included the event timestamp, source address, destination information, protocol, alert category, frequency, and source document identifier. Four alert groups were used in the evaluation: denial of service, brute force, exploit, and suspicious scan.

Elasticsearch stored the input alert documents, while SQLite maintained the durable queue, incident states, operational policies, whitelist and watchlist records, response rules, and audit information. The Core Orchestrator performed alert retrieval, normalization, duplicate prevention, deterministic classification, routing, and state management. Contextual analysis was provided by a local language model through Ollama, supported by a versioned Static RAG knowledge base and controlled by the Validator, Guardrail, and Output Sanitizer.

The Response Engine operated in laboratory modes that included dry run and controlled enforcement. Firewall related tests used isolated rules with time limited enforcement and rollback support. Model requests were subject to timeout, retry, token, and concurrency limits.

The source code retained llama3.2:3b only as a default value when no runtime model was provided. The official runtime configuration used for all LLM-related results reported in this paper overrode that default with qwen2.5:1.5b, served locally through Ollama 0.24.0. The same runtime model was used in the Validator and Guardrail evaluation, the eight-alert contention experiment, and the thirty latency measurements. The model tag, prompt and validator identifiers, knowledge base version, runtime parameters, and experimental evidence were frozen with the evaluated release. The study does not claim that the selected model is superior to other local language models.
The locked experimental configuration is summarized in
Table~\ref{tab:config}.
\begin{table}[!htbp]
\caption{LOCKED EXPERIMENTAL CONFIGURATION}\pdfbookmark[2]{Table \thetable: Locked experimental configuration}{bookmark:tab:config}
\label{tab:config}
\centering\scriptsize
\renewcommand{\arraystretch}{0.70}
\setlength{\tabcolsep}{2pt}
\begin{tabularx}{\columnwidth}{@{}L{0.34\columnwidth}X@{}}
\toprule
\textbf{Item} & \textbf{Configuration}\\
\midrule
Processor & Intel Core i5-12450H\\
Memory & 16 GB DDR4\\
Host environment & Windows 11 and Ubuntu on WSL\\
Elasticsearch & Version 8.12.0\\
Ollama & Version 0.24.0\\
Runtime model & qwen2.5:1.5b\\
Core workers & 4\\
LLM concurrency & 1\\
\bottomrule
\end{tabularx}
\end{table}

\subsection{Evaluation Scenarios}

The evaluation was organized into six groups. The first group examined deterministic classification and playbook routing at normal and boundary input values. The second group evaluated whitelist protection, safety checks, watchlist behavior, and controlled escalation. The third group examined human approval, duplicate decisions, dry run processing, response execution, time limited rules, and rollback.

The fourth group evaluated the local language model processing chain, including output parsing, field validation, guardrail replacement, output sanitization, and safe fallback behavior. The fifth group examined durable queue behavior, including duplicate prevention, retry, restart recovery, and a queue only stress experiment. The final group measured processing latency and evaluated contention for the local inference resource.

Additional hardening scenarios examined the recovery of incidents interrupted in the executing state. The recovery mechanism compared the stored incident state with the observed firewall state and did not automatically replay the response playbook.

\subsection{Evaluation Metrics and Pass Criteria}

Functional scenarios were evaluated using predefined expected states and pass or fail criteria. The principal observations included the assigned severity, selected playbook, queue state, incident state, approval state, execution state, language model status, firewall rule count, duplicate records, and audit information.

The baseline evaluation included ninety six functional and integration tests and thirty processing latency measurements. Additional experiments included five Output Sanitizer scenarios, two interrupted incident recovery scenarios, a queue only stress experiment with one hundred events, and an inference contention experiment with eight alerts.

The queue stress experiment was considered successful when all inserted events reached the completed state without duplicate event identifiers, duplicate incidents, new failed tasks, stuck executing incidents, or unintended firewall rules. The inference contention experiment evaluated semaphore acquisition and release counts, the maximum number of simultaneously active model requests, waiting time, task completion, duplicate records, and firewall effects.  The source code, experimental configuration, and test artifacts are publicly
available at \url{https://github.com/hhlam-gtvt/post-alert-incident-orchestration.git}.

\section{RESULTS AND DISCUSSION}\label{sec:results}

\subsection{Functional and Safety Results}

The baseline evaluation completed all predefined functional and integration scenarios with the expected states. Deterministic severity and playbook decisions remained independent of model output, safety controls prevented unsafe enforcement, and administrative decisions remained distinguishable from technical execution results. Table~\ref{tab:baseline} summarizes the evaluated groups.

\begin{table}[!htbp]
\caption{BASELINE FUNCTIONAL AND MEASUREMENT RESULTS}\pdfbookmark[2]{Table \thetable: Baseline functional and measurement results}{bookmark:tab:baseline}
\label{tab:baseline}
\centering\scriptsize
\renewcommand{\arraystretch}{0.70}
\setlength{\tabcolsep}{2pt}
\begin{tabularx}{\columnwidth}{@{}L{0.34\columnwidth}L{0.16\columnwidth}X@{}}
\toprule
\textbf{Evaluation group} & \textbf{Samples} & \textbf{Result}\\
\midrule
Rule Engine & 30 & 30 of 30 matched the predefined matrix\\
Whitelist and Safety Gate & 12 & 12 of 12 expected\\
Watchlist and escalation & 8 & 8 of 8 expected\\
Human in the loop & 10 & 10 of 10 expected\\
Firewall lifecycle & 12 & 12 of 12 expected\\
LLM, Validator, and Guardrail & 12 & 12 of 12 safely processed\\
Durable Queue & 12 & 12 of 12 expected\\
Processing latency & 30 & 30 of 30 completed\\
\bottomrule
\end{tabularx}
\end{table}

\subsection{Recovery and Output Control Results}

Additional hardening tests confirmed separation between administrative decisions and technical execution, non-replay recovery, and output sanitization. Three administrative decision scenarios passed. In two interrupted-execution cases, the Core reconciled an existing firewall rule without replaying the playbook, while an unverifiable state was moved to SAFETY\_HOLD. The Output Sanitizer passed all five predefined scenarios by removing or normalizing unsafe content as specified in the test criteria. Table~\ref{tab:hardening} summarizes these results.

Across three real Ollama calls in the baseline Validator evaluation, ten of thirteen required advisory fields were retained after validation, corresponding to 76.92\%. This ratio measures compliance with the implemented schema and grounding rules rather than semantic accuracy or intrusion-detection performance.

\begin{table}[!htbp]
\caption{ADDITIONAL HARDENING RESULTS}\pdfbookmark[2]{Table \thetable: Additional hardening results}{bookmark:tab:hardening}
\label{tab:hardening}
\centering\scriptsize
\renewcommand{\arraystretch}{0.70}
\setlength{\tabcolsep}{2pt}
\begin{tabularx}{\columnwidth}{@{}L{0.46\columnwidth}L{0.20\columnwidth}X@{}}
\toprule
\textbf{Evaluation} & \textbf{Scope} & \textbf{Result}\\
\midrule
Administrative decision handling & 3 scenarios & 3 of 3 passed\\
Interrupted execution recovery & 2 scenarios & 2 of 2 passed\\
Output Sanitizer & 5 scenarios & 5 of 5 passed\\
Playbook replay during recovery & 2 scenarios & 0 of 2 scenarios\\
\bottomrule
\end{tabularx}
\end{table}

\subsection{Queue Stress and LLM Contention Results}

The resource experiments evaluated two different processing paths. The
queue-only test completed 100 whitelisted events in 14.005 s without
invoking model inference or firewall execution, corresponding to an
isolated queue throughput of 7.14 events/s. In the LLM contention
experiment, eight medium-risk alerts were processed with four workers
and \texttt{LLM\_CONCURRENCY=1}. All tasks completed, the maximum
number of active model requests remained one, the maximum semaphore
wait was 117.524 s, and total experiment time was 307.137 s. No
duplicate tasks, failed records, lost incidents, or unintended firewall
rules were produced. Table~\ref{tab:resource} summarizes the measurements.

\begin{table}[!htbp]
\caption{RESOURCE EVALUATION RESULTS}\pdfbookmark[2]{Table \thetable: Resource evaluation results}{bookmark:tab:resource}
\label{tab:resource}
\centering\scriptsize
\renewcommand{\arraystretch}{0.70}
\setlength{\tabcolsep}{2pt}
\begin{tabularx}{\columnwidth}{@{}X L{0.25\columnwidth} L{0.25\columnwidth}@{}}
\toprule
\textbf{Metric} & \textbf{Queue only stress} & \textbf{LLM contention}\\
\midrule
Input alerts & 100 & 8\\
Completed tasks & 100 & 8\\
New failed tasks & 0 & 0\\
Duplicate queue tasks & 0 & 0\\
Duplicate incidents & 0 & 0\\
Firewall rules created & 0 & 0\\
Observed time & 14.005 s & 307.137 s\\
Queue-only observed throughput & 7.14 events per second & Not used\\
Maximum active LLM requests & Not applicable & 1\\
Maximum semaphore wait & Not applicable & 117.524 s\\
\bottomrule
\end{tabularx}
\end{table}

\subsection{Processing Latency}

\begin{figure}[!htbp]
\centering
\includegraphics[width=0.52\columnwidth,height=0.11\textheight,keepaspectratio]{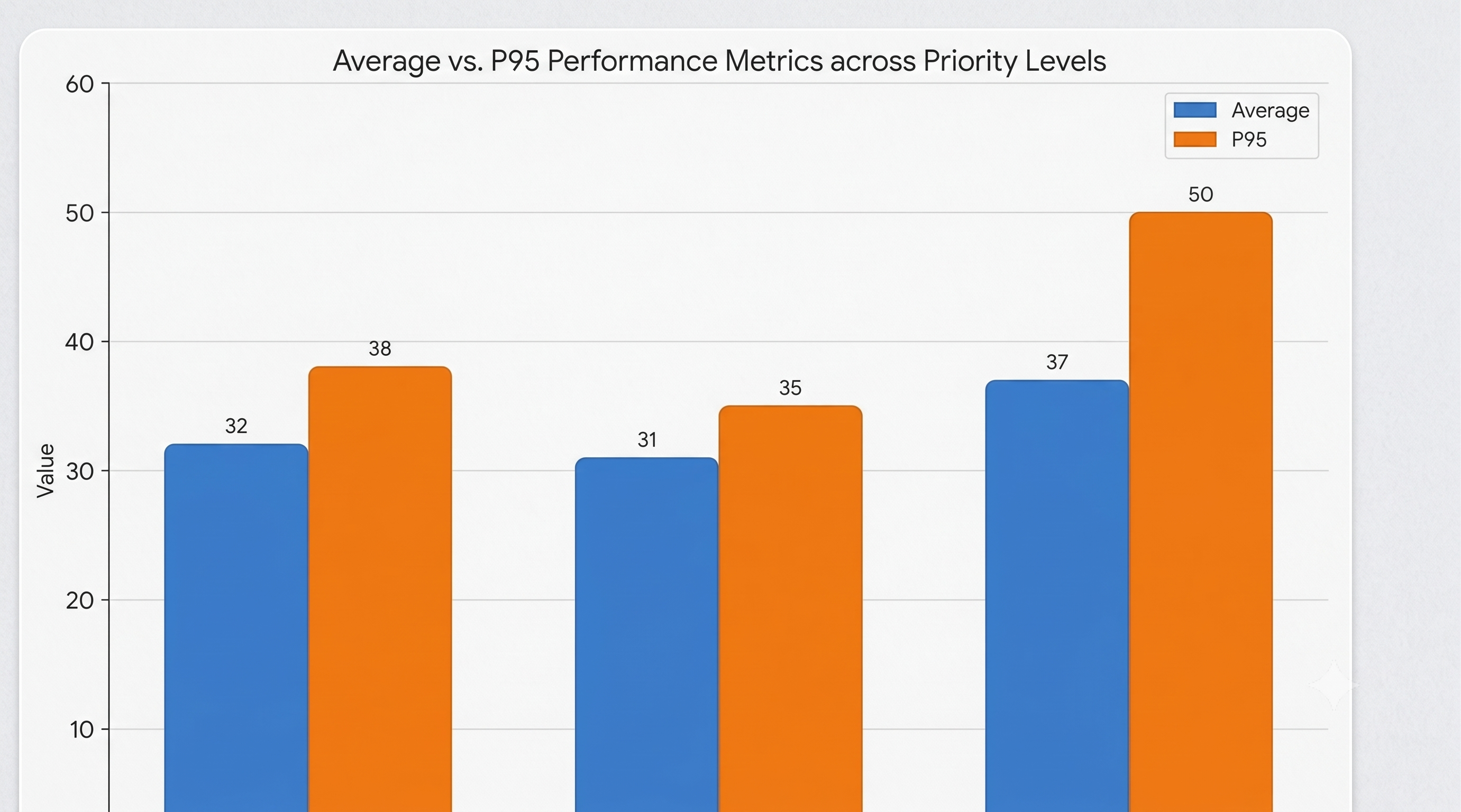}
\caption{Average and ninety fifth percentile processing latency by operating mode.}\pdfbookmark[2]{Figure \thefigure: Processing latency by operating mode}{bookmark:fig:latency}
\label{fig:latency}
\end{figure}

Table~\ref{tab:latency} and Fig.~\ref{fig:latency} summarize thirty
sequential measurements across the HIGH\_POST, MEDIUM\_PRE, and
LOW\_MONITOR modes using qwen2.5:1.5b. Mean processing times were approximately 32 s, 31 s, and 37 s, with corresponding P95 values of 38 s, 35 s, and 50 s. Across all samples, the overall mean was approximately 33 s and P95 was 45 s. Local model inference dominated the measured latency, while the Rule Engine, Validator, and Response Engine contributed only a small fraction of the total time.

\begin{table}[!htbp]
\caption{POST-ALERT PROCESSING TIME BY ANALYSIS MODE}\pdfbookmark[2]{Table \thetable: Post-alert processing time by analysis mode}{bookmark:tab:latency}
\label{tab:latency}
\centering\scriptsize
\renewcommand{\arraystretch}{1.0}
\setlength{\tabcolsep}{2pt}
\begin{tabular}{@{}lccc@{}}
\toprule
\textbf{Mode} & \textbf{Samples} & \textbf{Mean} & \textbf{P95}\\
\midrule
HIGH\_POST & 10 & 32 s & 38 s\\
MEDIUM\_PRE & 10 & 31 s & 35 s\\
LOW\_MONITOR & 10 & 37 s & 50 s\\
Overall & 30 & 33 s & 45 s\\
\bottomrule
\end{tabular}
\end{table}

Values are rounded from the recorded runtime measurements.

The experiments evaluate the operational integration and control of the local language model within the post-alert workflow rather than attempting to demonstrate that language-model-based analysis is superior to deterministic processing. The model is an advisory component and is not required for severity determination, playbook selection, authorization, or technical enforcement.

\subsection{Discussion}

The results indicate that deterministic processing should remain the default path for alert categories already represented by predefined rules and versioned Static RAG records. Under the evaluated configuration, mandatory local LLM invocation did not improve the automatic validity metrics and introduced substantial inference latency. The 100\% Guardrail intervention rate further shows that model output required systematic control before use.

This finding does not imply that language models are unsuitable for incident analysis. They may remain useful for unfamiliar alert categories, fragmented evidence, cross-incident synthesis, or analyst-facing explanation. The evaluated cases favored deterministic processing because they contained structured evidence and predefined knowledge records.

The queue and contention experiments also show that local model inference was the principal resource bottleneck. Increasing worker count alone would therefore not improve end-to-end performance without additional inference capacity. These findings apply only to the evaluated post-alert laboratory scenarios and do not establish intrusion-detection accuracy, production throughput, or real-time operational performance.

\section{LIMITATIONS}\label{sec:limitations}

The evaluation is limited to the post-alert processing path after an alert has been stored in Elasticsearch. It does not include packet capture, attack-traffic generation, intrusion-detection accuracy, or complete Logstash pipeline performance. The prototype was evaluated on one physical computer using local processes, containers, Windows Subsystem for Linux, SQLite, and a local Ollama service; therefore, the results do not demonstrate distributed coordination, high availability, horizontal scalability, or sustained production throughput.

The queue-only experiment intentionally bypassed language-model inference and firewall execution, while the contention and latency experiments showed that local inference was the main processing bottleneck. Rule Engine thresholds and escalation conditions were predefined for controlled evaluation rather than derived from long-term institutional traffic. Consequently, successful boundary testing demonstrates conformance to the implemented decision matrix but not the operational optimality of those thresholds.

The reported language-model results are specific to qwen2.5:1.5b, the evaluated prompt, Validator and Guardrail versions, Static RAG knowledge base, and hardware configuration. Static RAG uses manually maintained exact-category mappings, which improve reproducibility but limit coverage of unfamiliar or composite alerts. The ablation used ten synthetic cases repeated three times, and its automatic metrics evaluate conformity with predefined scoring rules rather than general semantic correctness or analyst usefulness.

Finally, the Output Sanitizer was evaluated with five predefined scenarios, and no long-term operational pilot with security administrators was conducted. The study therefore does not establish protection against every unsafe model output, nor does it measure analyst workload reduction, false-response rates, service interruption, user impact, or economic benefit.

\section{CONCLUSION}\label{sec:conclusion}
This paper presented a controlled post-alert incident orchestration and response subsystem for an educational information system. The proposed architecture separates deterministic classification, contextual analysis, administrative approval, and technical execution. The Rule Engine remains responsible for severity and playbook selection, while Static RAG and a local large language model provide advisory content under validation, Guardrail, Output Sanitizer, and Safe Fallback controls. Whitelist verification, durable processing, time-limited enforcement, rollback, recovery, and audit recording were integrated to preserve safety and traceability throughout the response workflow.

The experimental evaluation showed that the implemented components followed their predefined behavior under controlled laboratory conditions. The baseline evaluation included ninety-six functional and integration tests and thirty post-alert processing-time measurements. The Rule Engine, Whitelist and Safety Gate, Watchlist, human approval, firewall lifecycle, language model control chain, and Durable Queue produced the expected states in the evaluated scenarios. Additional hardening experiments showed that interrupted incidents were recovered without replaying their playbooks, and the Output Sanitizer passed all five defined scenarios.

The queue-only experiment completed all one hundred events without duplicate records, new failed tasks, stuck incidents, or unintended firewall rules. The language model contention experiment completed all eight alerts while maintaining the configured limit of one active inference request. However, local model inference remained the principal processing bottleneck, with an average post-alert processing time of approximately thirty-three seconds in the sequential latency experiment.

The analysis-layer ablation provided an additional boundary finding. The deterministic Rule Engine and Static RAG configuration met all automated validity criteria in all thirty observations, while the configuration using Static RAG, qwen2.5:1.5b, and Validator/Guardrail met all automated validity criteria in eighteen of thirty observations and required Guardrail intervention in every observation. Local model inference added an average analysis time of 37.638 seconds. Under the evaluated model, prompt, knowledge base, synthetic cases, and automatic scoring rules, mandatory language model invocation did not provide an advantage over the deterministic baseline for alert categories already represented by predefined rules and knowledge records.

These results support using deterministic processing as the default path and invoking the language model selectively when additional contextual interpretation is required. The findings demonstrate functional correctness, traceability, and controlled behavior within the evaluated scenarios, but they do not demonstrate intrusion detection accuracy, complete ELK pipeline performance, real-time response, or production readiness. Future work will integrate alerts from Suricata or another intrusion detection source, establish traffic baselines for threshold adjustment, and conduct a controlled pilot in an isolated laboratory or virtual local area network. Further evaluation should include longer workloads, unfamiliar alert categories, independent assessment by security analysts, alternative local language models, different hardware configurations, and a distributed queue and storage architecture.
\section*{ACKNOWLEDGMENT} 
Khuong Nguyen-An acknowledges the support of Ho Chi Minh City University of Technology (HCMUT), VNU-HCM, for his study.
\bibliographystyle{IEEEtran}
\bibliography{REFERENCES}

\end{document}